\documentclass{ws-ijmpd}
\newcommand{\PRL}{Phys. Rev. Lett. }
\newcommand{\PR}{Phys. Rev. }
\newcommand{\PL}{Phys. Lett. }
\newcommand{\CQG}{Class. Quantum Grav. }
\newcommand{\APJ}{Astrophys. J. }
\newcommand{\NP}{Nucl. Phys. }
\newcommand{\MNRAS}{Mon. Not. Roy. Astron. Soc. }
\newcommand{\be}{\begin{equation}}
\newcommand{\ee}{\end{equation}}
\begin{document}
\title{Constraints on a new alternative model to dark energy}
\author{Yungui Gong}
\address{Institute of Applied Physics and College of
Electronic Engineering, Chongqing University of Post and
Communication, Chongqing 400065, China\\ gongyg@cqupt.edu.cn}
\author{Xi-Ming Chen}
\address{College of Electronic Engineering, Chongqing University of Post and
Communication, Chongqing 400065, China} \maketitle
\begin{abstract}
The recent type Ia supernova data suggest that the Universe is
accelerating now and decelerated in recent past. This may provide
the evidence that the standard Friedmann equation needs to be
modified. We analyze in detail a new model in the context of
modified Friedmann equation using the supernova data published by
the High-$z$ Supernova Search Team and the Supernova Cosmology
Project. The new model explains recent acceleration and past
deceleration. Furthermore, the acceleration of the expansion of
the Universe is almost zero in the future.
\end{abstract}

\parindent=4ex
\section{Introduction}
The type Ia supernova (SN Ia) observations support that the
expansion of the Universe is speeding up rather than slowing down
\cite{sp97,tonry,riess04}. A spatially flat universe is favored by
the measurements of the anisotropy of the cosmic microwave
background (CMB) \cite{pdb00,bennett03}. The observation of type
Ia supernova SN 1997ff at $z\sim 1.7$ also provides the evidence
that the Universe is in the acceleration phase and was in the
deceleration phase in recent past \cite{agr,mstagr}. The
transition from the deceleration phase to the acceleration phase
happens around the redshift $z_{\rm T}\sim 0.4$
\cite{mstagr,daly}. The simplest way of explaining the
acceleration phenomenon is through the cosmological constant with
the equation of state parameter $\omega_{\rm \Lambda}=-1$. One
easily generalizes the cosmological constant model to dynamical
cosmological constant models, such as the dark energy model with
negative constant equation of state parameter $-1\le \omega_{\rm
Q}<-1/3$ and the holographic dark energy models
\cite{holo,kaloper}. In general, a scalar field $Q$ called
quintessence field that slowly evolves down its potential $V(Q)$
takes the role of a dynamical cosmological constant
\cite{quint,johri,aasss,uavs}. The energy density of the
quintessence field must remain very small compared with radiation
and matter at early epoches and evolves in a way that it started
to dominate the Universe around the redshift $0.4$. If we remove
the null energy condition restriction $\omega_{\rm Q}\ge -1$ to
allow supernegative $\omega_{\rm Q}<-1$, then the phantom energy
models arise \cite{phantom}. More exotic equation of state is also
possible, such as the Chaplygin gas model with the equation of
state $p=-A/\rho$ and the generalized Chaplygin gas models with
the equation of state $p=-A/\rho^\alpha$ \cite{chaply}. There are
fields other than the quintessence field, such as tachyon field,
as dark energy \cite{tachyon}.

Although there are many dark energy models proposed in the
literature, the nature of dark energy is still unknown. Therefore
it is also possible that the observations show a sign of the
breakdown of the standard cosmology. In brane cosmology, our
observable universe is a 3-brane which is located at the $Z_2$
symmetry fixed plane of a five dimensional Ads (Anti-de Sitter)
spacetime with the extra dimension to be an orbifold $S_1/Z_2$.
Bin\'etruy, Deffayet and Langlois showed that the Friedmann
equation takes the unconventional form $H^2\sim \rho +\rho^2$ in
brane cosmology \cite{brane}. In addition, Chung and Freese argued
that almost any relationship between $H$ and $\rho$ was possible
with the help of extra dimension \cite{chung}. Along this line of
reasoning, the usual Friedmann equation $H^2=8\pi G\rho/3$ is
modified to a general form $H^2=g(\rho)$
\cite{freese02,dvali,sen03,zhu03,wang,gong03a,gong03}. In these
models, no exotic matter form is needed and the Universe is
composed of the ordinary matter and radiation only. The function
$g(x)$ satisfies the following conditions: (1) $g(x_0)=1$; (2)
$g(x)\approx x$ when $z\gg 1$; (3) $g(x_0)>3x_0g'(x_0)/2$, where
$x=\Omega_{\rm m0}(1+z)^3+\Omega_{\rm r0}(1+z)^4$,
$x_0=\Omega_{\rm m0}+\Omega_{\rm r0}\approx \Omega_{\rm m0}$ and
$\Omega_{\rm r0}=8.40\times 10^{-5}$ \cite{bennett03}. In
Cardassian model, $g(x)=x+Bx^n$. In generalized Cardassian model,
$g(x)=x[1+Bx^{\alpha(n-1)}]^{1/\alpha}$ \cite{freese02}. For
Chaplygin gas like model, $g(x)=x+(A'+B'x^\beta)^{1/\beta}$
\cite{gong03a}. In Ref. \refcite{dvali}, Friedmann equation is
modified to be $H^2+H^\alpha\sim \rho$ based on the idea of
embedding a three brane into a five dimensional flat spacetime.
The modified model can be mapped into dark energy model. For
example, $g(x)=x$ for standard cosmology; $g(x)=x+\Lambda$ for
$\Lambda$CDM model; $g(x)=x+Bx^{1+\omega_{\rm Q}}$ for dark energy
models with constant equation of state parameter $\omega_{\rm Q}$.
In this paper, we discuss the new model proposed in Ref.
\refcite{gong03} in detail. We first derive the model with the
help of extra dimension, then we use different SN Ia data sets to
fit the model.

\section{Model Setup}
Follow Chung and Freese \cite{chung}, the new model motivated by
extra dimensions can be derived from the action
\begin{equation} S_5={-1\over 2\kappa_5}\int
d^5x\sqrt{\mathcal{G}}\,\mathbb{R}+S_{\rm orb}+S_{\rm
boundary},\end{equation} where $\mathbb{R}$ is the Ricci scalar in
five dimensions, $\mathcal{G}$ is the five dimensional metric
determinant, $\kappa_5=1/m^{3/2}_{pl,5}$ is the five dimensional
Newton's constant, $S_{\rm orb}$ is the orbifold or the bulk
action, and $S_{\rm boundary}$ represents the boundary or the
orbifold fixed plane action. In this setup, our universe is a
3-brane residing at the boundary of the fifth orbifold $S_1/Z_2$.
Varying the above action with respect to the metric, we get the
five dimensional Einstein equation
\begin{equation}
\label{5deq}
G_{MN}=\mathbb{R}_{MN}-\frac{1}{2}g_{MN}\mathbb{R}=\kappa^2_5
T_{MN}+\kappa^2_5 t_{MN}\delta(\Sigma),
\end{equation}
where $T_{MN}$ is the energy momentum tensor of the bulk and
$t_{MN}$ is the energy momentum tensor of the boundary.

To get cosmological solutions, we take the general five
dimensional metric for an isotropic and homogeneous flat brane
universe embedded in five dimensions as
\begin{equation}
ds^2=-N^2(t,y)dt^2+a^2(t,y)\delta_{ij}dx^idx^j+B^2(t,y)dy^2.
\end{equation}
The components of the five dimensional Einstein tensor are
\begin{equation}
\label{g00cop}
 G_{00}={N^2\over B^2}\left[-3{a''\over a}+3{a'\over
a} {B'\over B}-3\left({a'\over a}\right)^2\right]+3\left(
{\dot{a}\over a}\right)^2+3{\dot{a}\over a}{\dot{B}\over B},
\end{equation}
\begin{equation}
\label{g55cop}
 G_{55}=3{a'\over a}\left({a'\over a}+{N'\over
N}\right) +3{B^2\over N^2}\left[{\dot{a}\over a}{\dot{N}\over N}-
\left({\dot{a}\over a}\right)^2-{\ddot{a}\over a}\right],
\end{equation}
\begin{equation}
\label{g05cop}
 G_{05}=3{a'\over a}{\dot{B}\over B}+3{N'\over
N}{\dot{a}\over a} -3{\dot{a}'\over a},
\end{equation}
\begin{eqnarray}
\label{giicop}
 G_{ii}=&{a^2\over B^2}\left[{N''\over N}+2{a''\over
a} +\left({a'\over a}\right)^2+2{a'\over a}{N'\over N} -2{a'\over
a}{B'\over B}-{N'\over N}{B'\over B}\right]\nonumber\\
 &+{a^2\over
N^2}\left[-{\ddot{B}\over B}-2{\ddot{a}\over a}
-\left({\dot{a}\over a}\right)^2+2{\dot{N}\over N} {\dot{a}\over
a}-2{\dot{B}\over B}{\dot{a}\over a}+ {\dot{N}\over
N}{\dot{B}\over B}\right],
\end{eqnarray}
where ${\dot a} \equiv da/dt$, ${\ddot a}\equiv d^2 a/dt^2$ and
$a'\equiv da/dy$. Assume that the matter in the visible brane
located at the fixed plane $y=0$ takes the form of the perfect
fluid, then we get $t^M_N=\delta(y) {\rm diag}(-\rho,p,p,p,0)/B$.
Evaluate Eq. (\ref{5deq}) for the $G_{00}$ component Eq.
(\ref{g00cop}) and the $G_{ii}$ component Eq. (\ref{giicop}) at
the boundary $y=0$, we get the junction conditions
\begin{equation}
\label{jc1} \frac{[a']}{a^0B^0}=-\frac{\kappa^2_5}{3}\rho,
\end{equation}
\begin{equation}
\label{jc2} \frac{[N']}{N^0B^0}=\frac{\kappa^2_5}{3}(2\rho+3p),
\end{equation}
where superscript 0 means that the variable takes value at $y=0$
and the junction $[a']=a'(t,0_+)-a'(t,0_-)$. The junction of the
$G_{05}$ component Eq. (\ref{g05cop}) gives the usual energy
conservation equation
$$\dot{\rho}+3\frac{\dot{a}^0}{a^0}(\rho+p)=0.$$
The average of the $G_{55}$ component Eq. (\ref{g55cop}) at $y=0$
together with Eqs. (\ref{5deq}), (\ref{jc1}) and (\ref{jc2}) gives
\begin{equation}
\frac{1}{(N^0)^2}\left(\frac{(\dot{a}^0)^2}{(a^0)^2}-\frac{\dot{a}^0}{a^0}\frac{\dot{N}^0}{N^0}
+\frac{\ddot{a}^0}{a^0}\right)=-\frac{\kappa^2_5}{3(B^0)^2}T_{55}-\frac{\kappa^4_5}{36}\rho(\rho+3p).
\end{equation}
In terms of the cosmic time $d\tau=N^0dt$, the above equation can
be rewritten as
\begin{equation}
\label{breq1}
 \left(\frac{a^0_{,\tau}}{a^0}\right)^2
+\frac{a^0_{,\tau\tau}}{a^0}=-\frac{\kappa^2_5}{3(B^0)^2}T_{55}-\frac{\kappa^4_5}{36}\rho(\rho+3p).
\end{equation}
Now if we fix the gauge $B^0=1$ and assume that $T_{55}$ is a
constant, then we get the four dimensional Friedmann equation
\begin{equation}
\label{breq2}
H^2=\left(\frac{a^0_{,\tau}}{a^0}\right)^2=\frac{\kappa^4_5}{36}\rho^2-\frac{\kappa^2_5}{6}T_{55}+\frac{\mathcal{C}}{(a^0)^4},
\end{equation}
where $\mathcal{C}$ is an integration constant. If our 3-brane is
embedded in a five dimensional Ads spacetime, then
$T_{55}=\Lambda_B$ and $\rho=T+\rho_b$, here $-\Lambda_B$ is the
five dimensional cosmological constant, $\rho_b$ is the matter
density in the brane and $T$ is the brane tension. If for some
reasons, we have fine-tuned $T^2=6\Lambda_B/\kappa^2_5$ and
$\mathcal{C}=0$, then from Eq. (\ref{breq2}), we get the familiar
brane cosmological equation
$$H^2=\frac{\kappa^4_5}{36}\rho_b^2+\frac{\kappa^4_5}{18}T\rho_b.$$

If the bulk energy momentum tensor takes the perfect fluid form
$T^M_N={\rm diag}(-\rho_B,p_B,p_B,p_B,-\rho_B)$, then we can get
the solutions to Eq. (\ref{5deq}) as follows:
\begin{equation}
\label{5dsol1} B(t,y)=1,
\end{equation}
\begin{equation}
\label{5dsol2} \frac{\dot{a}(t,y)}{N(t,y)}=\alpha(t),
\end{equation}
\begin{equation}
\label{5dsol3} a^2(t,y)=A(t)\cosh(\mu y)+B(t)\sinh(\mu |y|)+C(t),
\end{equation}
where $\mu=\sqrt{-2\kappa^2_5\rho_B/3}$ in the case that
$\rho_B<0$ and $C(t)=3\alpha^2(t)/(\kappa^2_5\rho_B)$, or
\begin{equation}
\label{5dsol3a} a^2(t,y)=A(t)\cos(\mu y)+B(t)\sin(\mu |y|)+C(t),
\end{equation}
where $\mu=\sqrt{2\kappa^2_5\rho_B/3}$ in the case that
$\rho_B>0$, or
\begin{equation}
\label{5dsol3b} a^2(t,y)=\alpha^2(t)y^2+D(t)|y|+E(t),
\end{equation}
in the case that $\rho_B=0$. Furthermore, we can fix the temporal
gauge such that $N^0=1$. The integration constants $A(t)$, $B(t)$,
$D(t)$, $E(t)$ and $\alpha(t)$ can be determined from the junction
condition Eqs. (\ref{jc1}) and (\ref{jc2}).

If the bulk energy momentum tensor takes other forms, Chung and
Freese argued that almost any relationship between $\rho$ and $H$
was possible in Ref. \refcite{chung}. To see this, we take the
matter source in our brane Universe to be dust, i.e.,
$\rho=\rho_b\propto (a^0)^{-3}$ and $p=p_b=0$. Then the junction
condition Eqs. (\ref{jc1}) and (\ref{jc2}) can be rewritten as
\begin{equation}
\label{jc3}
\frac{2[a']}{a^0B^0}=-\frac{[N']}{N^0B^0}=-\frac{2\kappa^2_5}{3}\rho_b.
\end{equation}
If we need a relation $\rho_b=\nu H^q$, then we shall look for the
solution to the following equation
\begin{equation}
\label{4drel1}
\frac{3[a']}{a^0B^0}=-\nu\kappa^2_5\left(\frac{\dot{a}^0}{N^0
a^0}\right)^q.
\end{equation}
The solutions to Eqs. (\ref{jc3}) and (\ref{4drel1}) are
\begin{equation}
N(t,y)=B(t,y)=\exp(\beta(t)|y|),
\end{equation}
\begin{equation}
a(t,y)=\left(\frac{t}{t_0}\right)^{q/3}\exp(-\beta(t)[F(|y|)-F(y=0)]/2),
\end{equation}
where $\beta(t)=\kappa^2_5\nu(q/3t)^q/3$, $F(y)$ is any smooth
function satisfying $F'(y=0)=1$ and $t_0$ is an integration
constant. Substitute the above solutions to Eq. (\ref{5deq}), we
can get the bulk energy momentum tensor.

\subsection{New Model}
In ref. \refcite{gong03}, we proposed a new Friedman equation
\begin{equation}
\label{neweq} H^2=\sigma\rho_b[1+\exp(-\gamma\rho_b)]^n.
\end{equation}
In this section, we show how to derive the model. Now we are
looking for the solutions to the full five dimensional Einstein
equation (\ref{5deq}) satisfying the boundary condition Eq.
(\ref{jc3}) and the following equation
\begin{equation}
\label{4drel4}
\frac{\dot{a}^0}{N^0a^0}=\sqrt{\sigma}\sqrt{\frac{-3[a']}{\kappa^2_5
a^0 B^0}}\left(1+\exp\left[\frac{3\gamma[a']}{\kappa^2_5 a^0
B^0}\right]\right)^{n/2}.
\end{equation}
The general solutions which satisfy Eqs. (\ref{jc3}) and
(\ref{4drel4}) are
\begin{equation}
N(t,y)=B(t,y)=\exp(\kappa^2_5\rho_b(t)|y|/3),
\end{equation}
\begin{equation}
a(t,y)=a^0(t)\exp(-\kappa^2_5\rho_b(t)[F(|y|)-F(y=0)]/6),
\end{equation}
where $F(y)$ is any smooth function satisfying $F'(y=0)=1$,
$\rho_b(t)\propto (a^0)^{-3}$ and $a^0(t)$ are the solutions to
Eq. (\ref{neweq}). Substitute the above solutions into Eq.
(\ref{5deq}), we can get the five dimensional energy momentum
tensor which provides the new model.

In general, the modified Friedmann equations for a spatially flat
universe are
\begin{equation}
\label{cosa} H^2=H_0^2g(x),\ee \be \label{cosb} {\ddot{a}\over
a}=H^2_0g(x) -{3H^2_0x\over 2}g'(x)\left({\rho+p\over
\rho}\right),\ee\be \label{cosc} \dot{\rho}+3H(\rho+p)=0, \ee
where $x=8\pi G\rho/3H^2_0$, $1+z=a_0/a$ is the redshift
parameter, a subscript 0 means the value of the variable at
present, $g(x)$ is a general function of $x$ and from now on
$g'(x)=dg(x)/dx$. Let $\Omega_{\rm m0}=8\pi G\rho_{\rm
m0}/3H^2_0$, then $x_0=\Omega_{\rm m0}$ and $x=\Omega_{\rm
m0}(1+z)^3$ during matter dominated era. For our new model, we
have
\begin{equation}
\label{mymodel} g(x)=x(1+e^{-\alpha
x})^n=x[1+(x^{-1/n}_0-1)^{x/x_0}]^n,\end{equation} where
$\alpha=-\ln(\Omega_{\rm m0}^{-1/n}-1)/\Omega_{\rm m0}>0$.

Dark energy models have been extensively studied in the
literature. Most researchers studied dark energy models by
constraining the dark energy equation of state parameter
$\omega_{\rm Q}$. To compare the modified model with the dark
energy model, we make the following identification
\begin{equation}
\label{wq} \omega_{\rm Q}={xg'(x)-g(x)\over g(x)-x}.
\end{equation}
The transition from deceleration to acceleration happens when the
deceleration parameter $q=-\ddot{a}/aH^2=0$. From equations
(\ref{cosa}) and (\ref{cosb}), the transition redshift $z_{\rm T}$
is given by
\begin{equation}
\label{trans} g[\Omega_{\rm m0}(1+z_{\rm T})^3]={3\over
2}\Omega_{\rm m0}(1+z_{\rm T})^3g'[\Omega_{\rm m0}(1+z_{\rm
T})^3].
\end{equation}

\section{Supernova fitting result}
The current age of the Universe is \begin{equation}
H_0t_0=H_0\int^{t_0}_0dt=\int^\infty_0{dz\over
(1+z)g^{1/2}[\Omega_{\rm m0}(1+z)^3]} \end{equation}

The luminosity distance $d_{\rm L}$ is defined as
\begin{equation}
\label{lumin} d_{\rm L}(z)=a_0(1+z)\int^{t_0}_t {c\,dt'\over
a(t')}={c(1+z)\over H_0}\int^z_0 g^{-1/2}[\Omega_{\rm
m0}(1+u)^3]du.
\end{equation}

The apparent magnitude redshift relation is
\begin{equation}
\label{magn}
 m(z)=M+5\log_{10}d_{\rm L}(z)
+25=\mathcal{M}+5\log_{10}\mathcal{D}_{\rm L}(z),
\end{equation}
where $\mathcal{D}_{\rm L}(z)=H_0d_{\rm L}(z)$ is the
``Hubble-constant-free" luminosity distance, $M$ is the absolute
peak magnitude and $\mathcal{M}=M-5\log_{10}H_0+25$.  The
parameters in our model are determined by minimizing
\begin{equation}
\label{lrmin} \chi^2=\sum_i{[m_{\rm obs}(z_i)-m(z_i)]^2\over
\sigma^2_i},
\end{equation}
where $\sigma_i$ is the total uncertainty in the observation. The
$\chi^2$-minimization procedure is based on MINUIT code
\cite{minuit}. The observational data used are: the 20 radio
galaxy and 78 supernova data by Daly and Djorgovski \cite{daly},
the 172 supernova data with $z>0.01$ and $A_v<0.5$ mag by Tonry
{\it et al.} and the 22 supernova data with $A_v<0.5$ mag by
Barris {\it et al.} \cite{tonry}, and the latest 157 supernova
data compiled by Riess {\it et al.} \cite{riess04} combined with
the CMB shift parameter $\mathcal{R}=\Omega_{\rm
m0}^{1/2}H_0d_{\rm L}(z_{\rm ls})/(c(1+z_{\rm ls}))=1.716\pm
0.062$ \cite{shift}, here $z_{\rm ls}=1089\pm 1$ \cite{bennett03}.
For the SN Ia data compiled by Tonry {\it et al.} and Barris {\it
et al.}, a 500 km/s uncertainty in quadrature to the redshift
errors is added to account for the velocity uncertainty and
$\chi^2$ is calculated by marginalizing over $H_0$. The
marginalization is done by integrating $e^{-\chi^2/2}$ over all
possible values of $H_0$. Since $H_0$ is appeared linearly in
$\chi^2$, the integration process is equivalent to minimize
$\chi^2$ over $H_0$.

From equation (\ref{mymodel}), it is easy to see that at high
redshift $z$, $x$ is very large and $g(x)\sim x$. Therefore the
standard model is recovered at early times. In the future,
$\exp(-\alpha x)\to 1$ and $g(x)\approx 2^n x$, so the Universe
will again evolve as $t^{2/3}$.  Substituting the new model
(\ref{mymodel}) into equation (\ref{cosb}), during the matter
domination we get
\begin{equation}
\label{qzrel} {\ddot{a}\over a H^2_0}=-{1\over 2}x(1+e^{-\alpha
x})^n+{3\over 2}n\alpha x^2(1+e^{-\alpha x})^{n-1}e^{-\alpha x}.
\end{equation}
As $x\gg 1$, we have $\ddot{a}/(a H^2_0)\rightarrow -x/2$. As
$x\rightarrow 0$, $\ddot{a}/(a H^2_0)\rightarrow 0$. In this
model, equation (\ref{trans}) becomes
\begin{equation}
1+(\Omega_{\rm m0}^{-1/n}-1)^{(1+z_{\rm T})^3}+3n\ln(\Omega_{\rm
m0}^{-1/n}-1) (\Omega_{\rm m0}^{-1/n}-1)^{(1+z_{\rm
T})^3}(1+z_{\rm T})^3=0
\end{equation}
If we compare this model with the dark energy model, from Eq.
(\ref{wq}), we get
$$\omega_{\rm Q0}={n(1-\Omega_{\rm m0}^{1/n})\ln(\Omega_{\rm m0}^{-1/n}-1)\over
\Omega_{\rm m0}(1-\Omega_{\rm m0})}.$$

In the fit, the range of parameter space for $n$ is $(0,\ 45]$ and
the range of parameter space for $\Omega_{\rm m0}$ is $(0,1]$. The
best fit results are summarized in table \ref{fittab}. In table
\ref{fittab}, we also list the transition redshift, the state of
equation parameter $\omega_{\rm Q0}$ of the equivalent dark energy
model and the age of universe using the best fit parameters. For
the last row in table \ref{fittab}, we fit the model to both the
157 gold sample supernova data and the CMB shift parameter.  In
this model, the Universe is older than that by the standard
cosmology. The contour plots to different data sample are shown in
Figs. \ref{dalyfit}-\ref{wqvsz}.
\begin{table}
\caption{Summary of the best fit parameters. In some cases, the
parameter reaches the limit of the range of the parameter space.}
\label{fittab}
\begin{center}
\begin{tabular}{|c|c|c|c|c|c|c|c|c|}
  \hline
 Data & \multicolumn{2}{|c|}{$\Omega_{\rm m0}$} & \multicolumn{2}{|c|}{$n$} & $\chi^2$&$\omega_{\rm Q0}$ & $z_{T}$&$H_0t_0$ \\
  \cline{2-5}
   \# & 68\% & 99\% & 68\% & 99\% &  & & & \\
  \hline
   98 & $0.33^{+0.19}_{-0.14}$ & $0.33^{+0.29}_{-0.15}$ & $2.46^{+0.31}_{-0.34}$ & $2.46_{-0.6}^{+12}$ &
  88.1&-2.27&0.76&0.97
  \\\hline
   194 &$0.54^{+0.06}_{-0.09}$ & $0.54^{+0.12}_{-0.09}$ & $3.34^{+7.21}_{-1.22}$ & $3.34_{-1.67}^{+108}$ &195.96&
  -3.62&0.30&0.86
  \\\hline
 158 &$0.46^{+0.09}_{-0.11}$ & $0.46^{+0.15}_{-0.21}$ & $2.60^{+4.70}_{-0.54}$ & $\ge 1.75$ &181.4&
  -2.85 &0.44&0.90
  \\\hline
\end{tabular}
\end{center}
\end{table}

\begin{figure}[htb]
\vspace{-0.1in}
\begin{center}
\epsfxsize=3.3in \epsffile{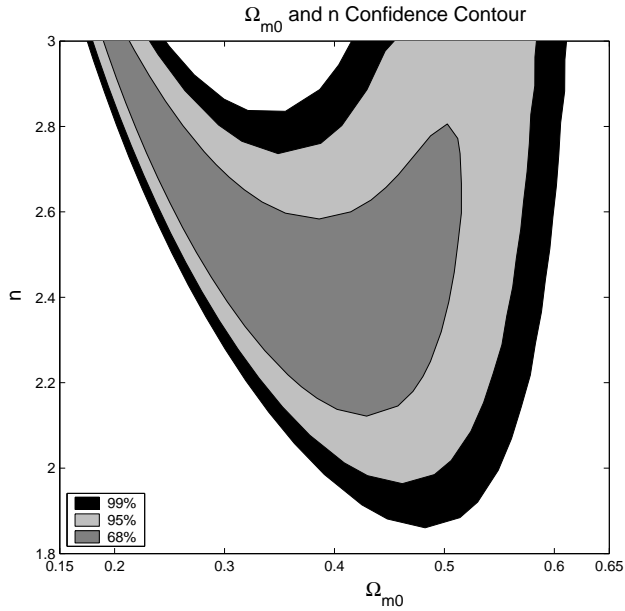}
\end{center}
\vspace{-0.2in} \caption{The $\Omega_{\rm m0}$ and $n$ contour
plot for the 20 radio galaxy and 78 supernova data by Daly and
Djorgovski} \label{dalyfit}
\end{figure}

\begin{figure}[htb]
\vspace{-0.1in}
\begin{center}
\epsfxsize=3.3in \epsffile{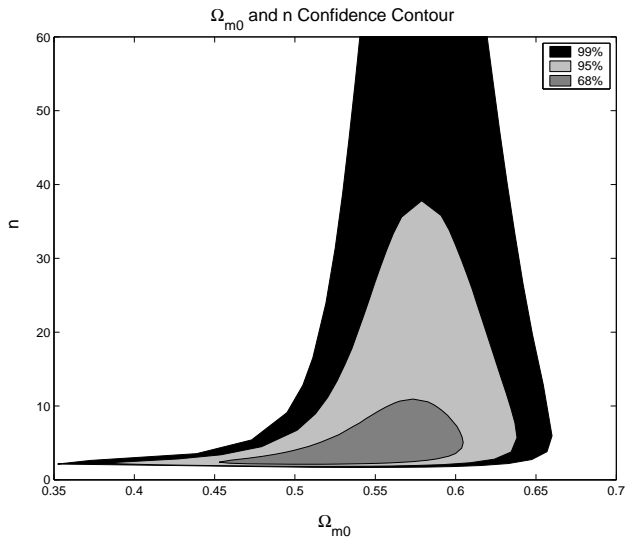}
\end{center}
\vspace{-0.2in} \caption{The $\Omega_{\rm m0}$ and $n$ contour
plot for the 194 combined supernova data by Tonry {\it et al}. and
Barris {\it al.}} \label{tonbarfit}
\end{figure}

\begin{figure}[htp]
\vspace{-0.1in}
\begin{center}
\epsfxsize=3.3in \epsffile{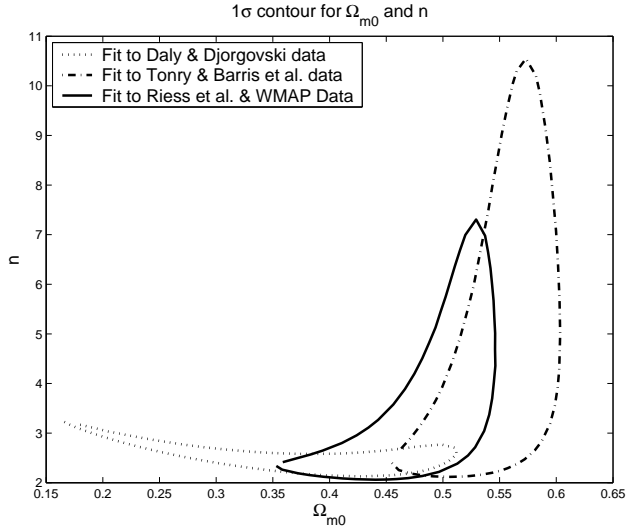}
\end{center}
\vspace{-0.2in} \caption{The $1\sigma$ $\Omega_{\rm m0}$ and $n$
contour plots} \label{combfit}
\end{figure}

\begin{figure}[htb]
\vspace{-0.1in}
\begin{center}
\epsfxsize=3.3in \epsffile{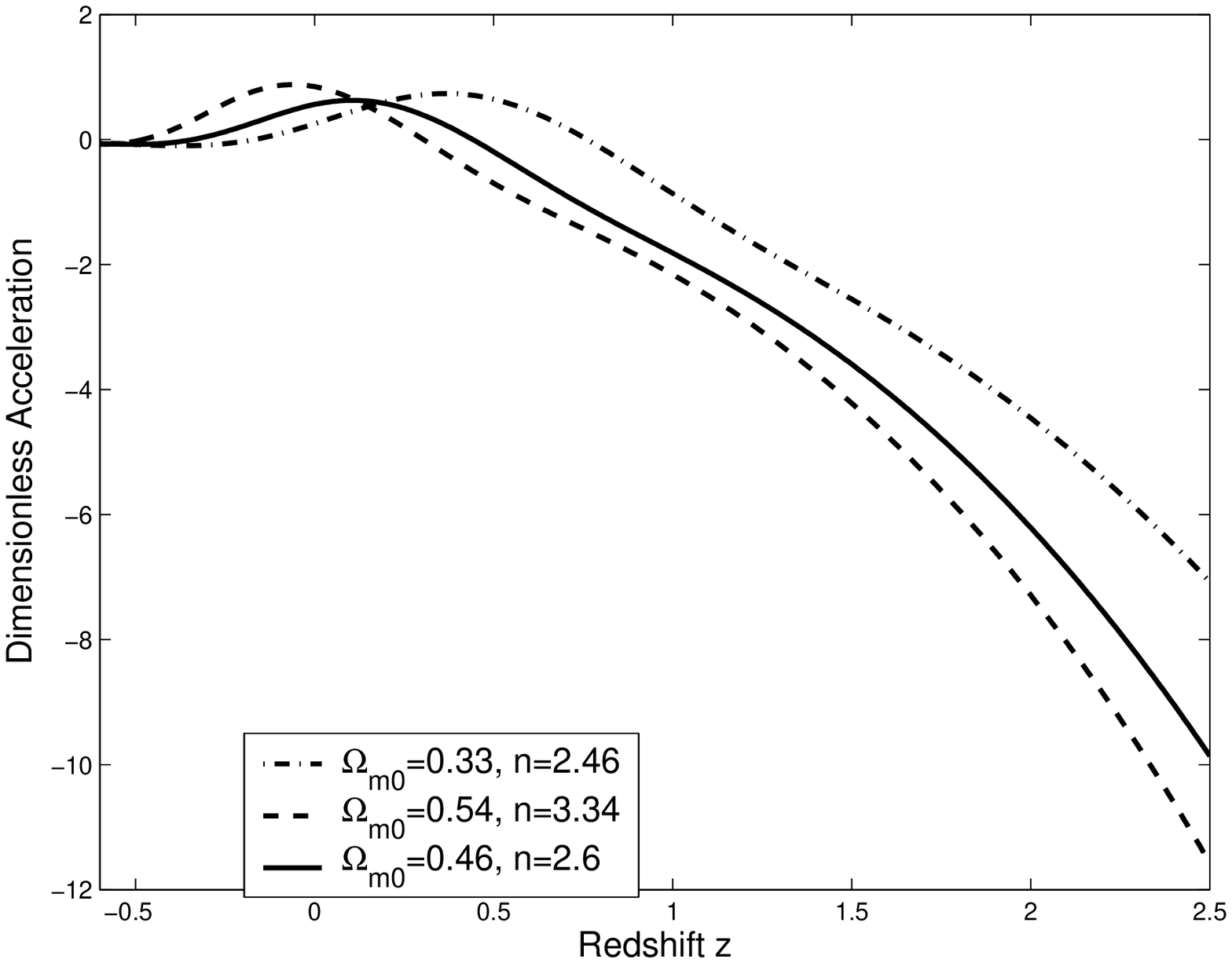}
\end{center}
\vspace{-0.2in} \caption{$\ddot{a}/aH^2_0$ versus the redshift $z$
for the best fit parameters listed in Table I.} \label{bestfit}
\end{figure}

In Fig. \ref{bestfit}, we plot the dimensionless acceleration
$\ddot{a}/aH^2_0$ against the redshift. It is clear that the model
have deceleration phase in recent past and acceleration phase at
present. In Fig. \ref{wqvsz}, the evolution of $\omega_{\rm Q}$ of
the equivalent dark energy model is shown for the best fit
parameters to the combined supernova and CMB data. Around $z\sim
2$, $g(x)\approx x$ and the Universe became indistinguishable from
the standard matter dominated universe.

\begin{figure}[htb]
\vspace{-0.1in}
\begin{center}
\epsfxsize=3.3in \epsffile{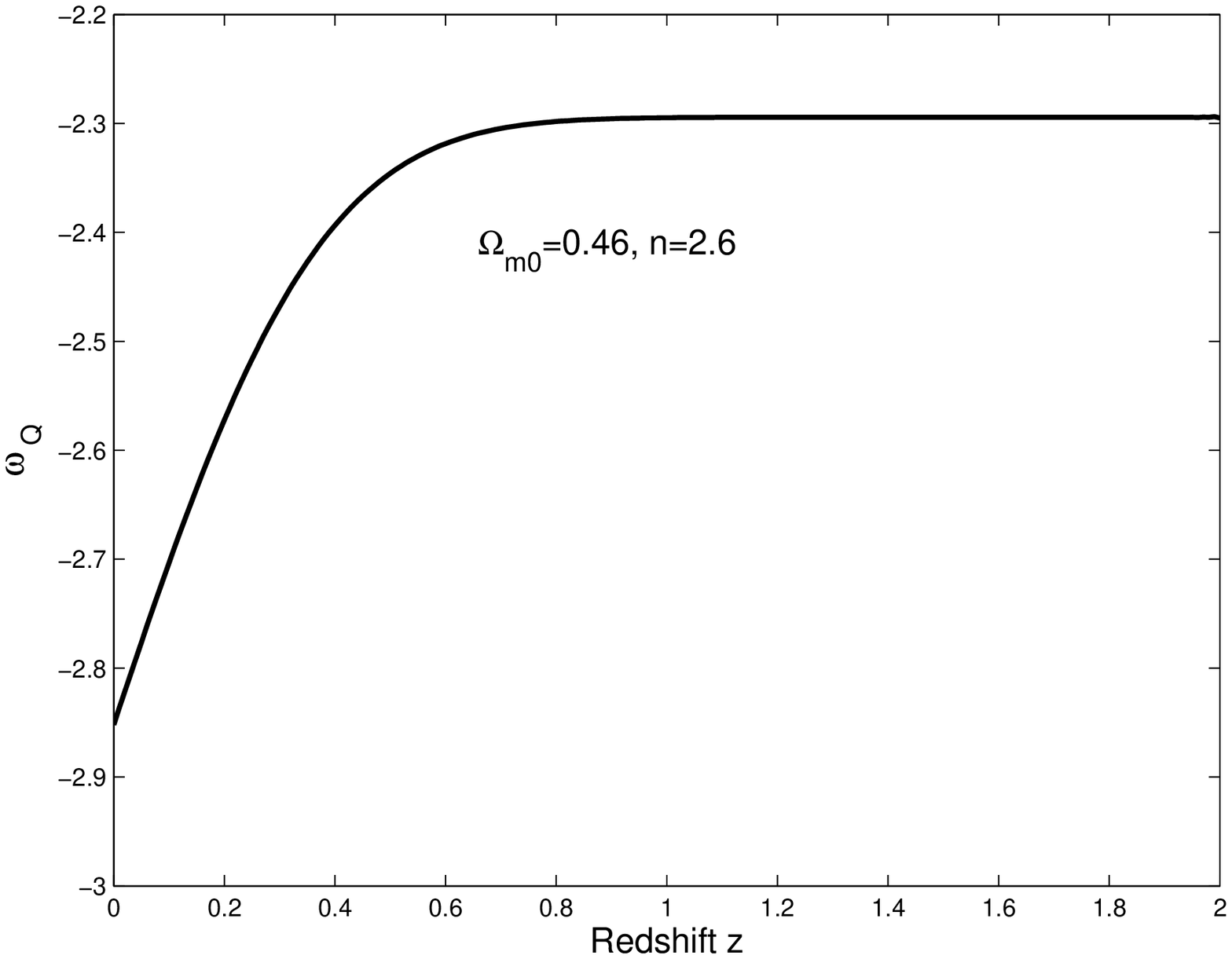}
\end{center}
\vspace{-0.2in} \caption{$\omega_{\rm Q}$ versus the redshift z
for the best fit parameters $\Omega_{\rm m0}=0.52$ and $n=2.8$.}
\label{wqvsz}
\end{figure}

As shown in Fig. \ref{bestfit}, the new model gives past
deceleration, current acceleration and future deceleration or near
zero acceleration. From theoretical point of view, future
deceleration is favored because it avoids the horizon problem. In
Ref. \refcite{kaloper}, the authors show that eternal accelerating
dark energy models prevent us from ever measuring inflationary
perturbations which originated before the ones currently
observable due to the growth of the comoving Hubble scale in the
future. The Universe in this new model is older than that in
standard cosmology as shown in table \ref{fittab}. Gong and Daly
and Djorgovski found the model independent result $z_{\rm T}\sim
0.4$ \cite{daly}. The result is consistent with ours as seen from
table \ref{fittab}. The somewhat model independent study by Alam,
Sahni, Saini and Starobinsky found that $\omega_{\rm Q}$ evolves
rapidly from zero at high $z$ to a strongly negative value at
present \cite{uavs}. This result is also consistent with ours as
shown in table \ref{fittab} and Fig. \ref{wqvsz}. In conclusion,
we proposed a new model which is consistent with current supernova
observations and avoids the future horizon problem. It is
difficult to discriminate this model from dark energy models with
current observations. Future supernova data from the dedicated
satellite telescope (such as Supernova/Acceleration Probe) is
necessary to differentiate these models.

\section*{acknowledgments}
The work is supported by Chongqing University of Post and
Telecommunication under grants A2003-54 and A2004-05.

\end{document}